# NMPC and Deep Learning-Based Vibration Control of Satellite Beam Antenna Dynamics Using PZT Actuators and Sensors

Sean Kalaycioglu[1,] Daniel Ding[2],
[1]Department of Aerospace Engineering, Toronto Metropolitan University, Toronto, ON, Canada-M5B 2K3
[2]UCC, ON, Canada, M4V 1W6

*Abstract*—This paper presents a novel approach for vibration control of satellite-based flexible beam-type antennas using Nonlinear Model Predictive Control (NMPC) and Deep Learning techniques. The developed control system leverages piezoelectric (PZT) actuators and sensors to manage the coupled attitude and structural dynamics of the satellite, improving precision and stability. We propose a detailed coupled dynamics model that integrates both satellite attitude and beam structural dynamics, considering the effects of PZT-based actuators. Through MATLAB/Simulink simulations, we demonstrate the effectiveness of the combined NMPC and Deep Learning framework in reducing structural vibrations, achieving faster response times, and enhancing overall control accuracy. The results indicate that the proposed system provides a robust solution for controlling flexible beam-type satellite antennas in space environments.
*Keywords*— NMPC, Deep Learning, Vibration Control, Piezoelectric (PZT) Actuators and Sensors, Structural Dynamics

## I. INTRODUCTION

In modern satellite systems, the increasing demand for higher data transmission rates and efficient communication capabilities has led to the use of flexible beam-type antennas. These antennas, while highly efficient, are prone to structural vibrations due to their flexibility, which can significantly degrade performance. Controlling these vibrations is critical, particularly when the dynamics of the satellite's attitude are coupled with the vibrations of the antenna. The integration of advanced control techniques such as Nonlinear Model Predictive Control (NMPC) and Deep Learning, along with the use of piezoelectric (PZT) actuators and sensors, represents a promising avenue for solving this challenge.

Several studies have tackled the issue of vibration control in satellite systems. Ji et al. (2024) proposed an adaptive fault-tolerant control framework for flexible satellites, reducing vibration through an event-triggered mechanism based on a PDE model ([Ji, 2024]). Similarly, Callipari et al. (2022) investigated Offset Piezoelectric Stack Actuators (OPSA) for enhanced damping in space structures, outperforming conventional piezoelectric patches ([Callipari, 2022]).

Active vibration control (AVC) has also been explored extensively. Smith and Johnson (2018) reviewed various AVC strategies for large flexible space structures, highlighting the use of PZT actuators to minimize vibrational disturbances [Smith, 2018]. Angeletti et al. (2020) introduced a distributed network of actuators and sensors, showing that optimal actuator placement significantly reduces deformations during attitude maneuvers ([Angeletti, 2020]). Similarly, Lee et al. (2020) applied AVC to large flexible satellite structures using PZT sensors and actuators [Lee, 2020]. Zhang et al. (2020) demonstrated adaptive vibration control using fixed-time prescribed performance control, achieving high precision during dynamic satellite operations [Zhang, 2020]. The integration of Deep Learning into control strategies has garnered significant attention. Qiu and Wang (2021) developed neural network models for adaptive control, demonstrating that deep learning can predict optimal control inputs based on past vibration data [Qiu, 2021]. Sun and Li (2021) further showed that combining deep learning with NMPC can optimize real-time control, leading to improved satellite stability [Sun, 2021]. Zhou et al. (2020) also emphasized the role of deep learning in optimizing control algorithms for large, flexible satellite structures [Zhou, 2020].

The role of PZT actuators and sensors has been studied extensively. Karami and Daneshmand (2019) explored the enhancement of attitude control and vibration mitigation through PZT actuators [Karami, 2019]. Abdo and Kaddouh (2021) examined the nonlinear behavior of PZT patches in vibration suppression, concluding that smart materials play a pivotal role in modern satellite control systems [Abdo, 2021]. Wang et al. (2021) proposed a hybrid system integrating PZT sensors with NMPC to improve vibration suppression in flexible satellite appendages [Wang, 2021]. Sharma et al. (2020) highlighted the potential of piezoelectric smart materials in creating self-regulating satellite structures capable of mitigating vibrations autonomously [Sharma, 2020].

Despite these advances, challenges remain in optimizing actuator placement and fully integrating NMPC with smart materials. Park and Shin (2021) called for more research into optimizing actuator placement for smart materials to fully exploit their capabilities in vibration control [Park, 2021]. Tang and Zhou (2019) emphasized the need for improved placement strategies for smart actuators to better manage vibrational modes in flexible structures [Tang, 2019].

Furthermore, deep learning presents a promising solution to many of these challenges. Zhao and Cheng (2020) discussed how deep learning could improve real-time control in rapidly changing space environments, where traditional models struggle [Zhao, 2020]. Zou and Li (2019) explored how reinforcement learning can be integrated with control systems to adapt to environmental uncertainties and improve vibration suppression during satellite operations [Zou, 2019]. Kumar and Nair (2019) reviewed optimization strategies for controlling vibrations in flexible space antennas, stressing the importance of adaptive control for mission success [Kumar, 2019].

Lee and Han (2021) demonstrated that predictive control algorithms integrated with machine learning can further stabilize satellite systems with flexible appendages [Lee, 2021].

In parallel, Li and Chen (2020) explored how NMPC integrated with PZT sensors can improve control precision in satellite applications, particularly in flexible structures [Li, 2020]. Anderson and Zhang (2020) modeled the coupled dynamics of smart actuators in flexible satellites, revealing the significant improvements in stability and performance achieved by integrating PZT actuators with NMPC [Anderson, 2020]. Wu and Feng (2022) highlighted how advanced PZT-based control systems can manage large-scale structures under real-time operational conditions [Wu, 2022].

Lastly, integration of advanced optimization strategies and smart actuator networks was examined by Park and Kim (2020), who showed that integrating PZT sensors into attitude control systems improves precision, especially during high-frequency disturbances [Park, 2020]. Zhao and Cheng (2020) further emphasized the role of deep learning in dynamically controlling flexible structures, providing solutions to challenges in space environments [Zhao, 2020]. Gao and Yu (2018) focused on finite element modeling to optimize flexible satellite structures, highlighting the significance of accurate structural simulations for control system design [Gao, 2018].

In conclusion, significant progress has been made in the development of control strategies for flexible satellite antennas. However, challenges remain, particularly in integrating NMPC, deep learning, and PZT actuators. This paper aims to address these gaps by proposing an integrated control system leveraging these advanced technologies for precise vibration suppression and satellite attitude stabilization.

This paper is organized as follows: In Section 2, a detailed introduction to the system is provided, covering the spacecraft's geometric configuration, including its flexible components and structural dynamics. Section 3 delves into the forces and moments generated by piezoelectric actuators. Section 4 outlines two innovative control approaches for vibration mitigation: the NMPC model and the NARX-based control method, explaining the theoretical foundations and implementation procedures for each. Section 5 discusses the results, comparing the performance of the proposed control techniques against conventional methods under different disturbance scenarios. Finally, Section 6 summarizes the key findings of the research, highlights their significance, and proposes potential practical applications.

## II. OVERVIEW OF THE SYTEM

Our research examines a rigid spacecraft equipped with $n$ slender, deformable appendages integrated with piezoelectric (PZT) control elements. Figure 1 illustrates this setup, showcasing a specific example in which two elastic beams are attached to the central body. The piezo-ceramic actuators and strain sensors, either bonded to or embedded within the spacecraft's appendages, are depicted in Figure 2.

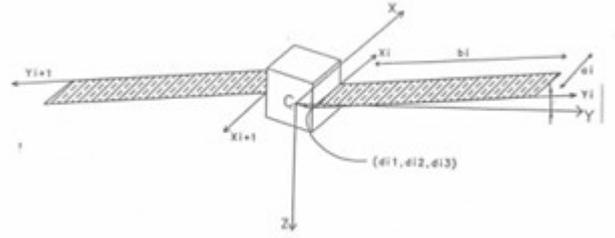

Fig. 1 Satellite with Flexible Beam-Type Appendages

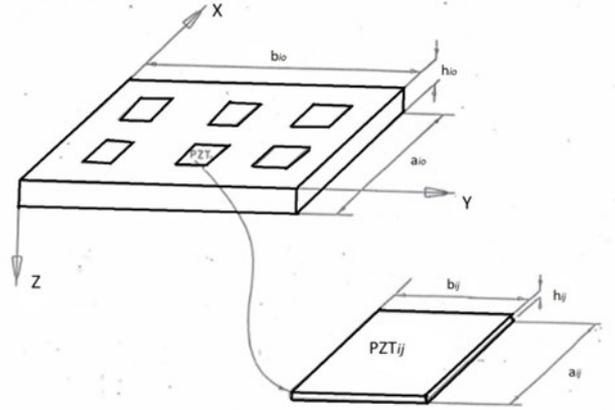

Fig. 2. Geometry of the beam and the piezo-ceramic actuators

The vibration equations for a beam-type structure are derived using the Lagrangian formulation. In this approach, the kinetic energy ($T$), potential energy ($V$) of the beam, and the generalized forces generated by the piezo-ceramic actuators are systematically calculated.

The transverse displacement of the beam, denoted as $u(y_i,t)$, occurs along the $z$-direction for the $i^{th}$ appendage. For ease of analysis, the elastic displacement $u$ is expressed as a series expansion, as follows:

$$u_i(y_i,t) = \sum_{s=1}^{n_s} W_{is}(t)\psi_s(y_i) = W^T \vec{\Psi}_i \qquad (1)$$

The spatial functions $\psi_s(y_i)$ constitute a complete orthogonal set that satisfies the structural boundary constraints. Increasing $n_s$ enhances the accuracy of the approximation, allowing convergence toward the true displacement field. In practical applications, selecting appropriate $\psi_s(y_i)$ facilitates accurate representation through truncated series expansions. The natural modes of fixed-free structural elements are particularly effective as basis functions for $\psi_s(y_i)$, as these eigen-solutions inherently satisfy both geometric and force boundary conditions while preserving orthogonality within the solution domain.

The kinetic energy of the beam, $T$ can be written as:

$$T = \frac{1}{2}\int_0^{b_i} (\dot{\vec{R}}_i)^T \cdot (\dot{\vec{R}}_i)\frac{m_i}{b_i}\,dy_i \qquad (2)$$

where $m_i$ is the mass density per unit length in the y-direction and $\dot{R}_i$ is the velocity of any point on the beam.

The strain energy stored in the beams can be expressed as:

$$U = \frac{1}{2}\sum_{i=1}^{n} D_i \int_0^{b_i}\left[\left(\frac{\partial^2 u_i}{\partial y_i^2}\right)^2\right] dy_i \quad (3)$$

where

$$D_i = \frac{E_{io}h_{io}^3}{12(1-\nu^2)}$$

The parameter $v$ is the Poisson ratio.

$D_i$ represents the flexural rigidity of the beam, associated with its transverse displacement along the z-axis. The governing equations for the beam's vibrations are derived using the Lagrangian formulation, as follows:

$$\frac{d}{dt}\left[\frac{\partial T}{\partial \dot{W}_{is}}\right] - \left[\frac{\partial T}{\partial W_{is}}\right] + \left[\frac{\partial U}{\partial W_{is}}\right] = \vec{F}_i \quad (4)$$

The index $s$ ranges from 1 to $n_r$, where $n_r$ represents the truncation order in the spectral decomposition for clamped-free (C-F) boundary conditions. The term $F_i$ corresponds to the generalized modal force components.

The generalized modal force components $F_i$ account for contributions from external disturbances, dissipative effects, and controlled actuation inputs. Consider a distributed environmental load $p(y_i,t)$ acting normal to the surface of the i-th flexible member, alongside discrete control forces $\tau_r$ applied perpendicularly at specific points $y_{ir}$, measured from the member's base. The resulting modal forcing function can be expressed as:

$$\vec{F}_i = \int_0^{b_{io}} p(y_i,t)\psi_j\, dy_i + \sum_{r=1}^{p} \tau_r \psi_j(y_{jr}) + \vec{F}_{ds} \quad (5)$$

The dissipative force component $F_{ds}$ is represented as a linear function of the generalized velocity capturing uncoupled viscous energy dissipation in the modal domain. By incorporating the derived energy expressions into the variational formulation and carrying out the necessary mathematical derivations, the following result is obtained:

$$\ddot{\vec{W}}_i + D_{wi}\dot{\vec{W}}_i + K_i \vec{W}_i = \vec{F}_i \quad (6)$$

which governs the vibrations of the beam, $D_{wi}$ is the damping matrix and is assumed to be diagonal. $F_i$ represents the generalized forces due to the piezo-ceramics actuator control forces $F_c$ and the other dynamics forces including the external disturbances $F_d$:

$$\vec{F}_i = \vec{F}_c + \vec{F}_d \quad (7)$$

The subsequent analysis focuses on evaluating $F_c$ specifically for piezo-ceramic actuators.

## III. GENERALIZED MOMENTS AND FORCES CREATED BY PIEZOCERAMICS ACTUATORS

The stress in the $PZT_{ij}$ can be written as:

$$\sigma_{ij} = E_{ij}(\epsilon_{io} + \epsilon_{ij}) \quad (8)$$

where $E_{ij}$ represents the modulus of elasticity, $\epsilon_{ij}$ is the strain in the $PZT_{ij}$ (caused by the applied voltage), and $\epsilon_{io}$ is the longitudinal strain in the beam due to force balance. The strain $\epsilon_{ij}$ can be characterized by:

$$\epsilon_{ij} = \frac{d_{31}V_{ij}(t)h_{ij}}{b_{ij}} \quad (9)$$

where $d_{31}$ is the PZT electric coefficient, $h_{ij}$ represents the thickness of $PZT_{ij}$, and $V_{ij}(t)$ denotes the voltage applied to $PZT_{ij}$.

Application of force equilibrium principles to the differential element yields:

$$E_{ij}h_{ij}b_{ij}(\epsilon_{ij}+\epsilon_{io}) + E_{io}h_{io}b_{io}\epsilon_{io} = 0 \quad (10)$$

Thus, the longitudinal strain $\epsilon_{io}$ is:

$$\epsilon_{io} = \frac{E_{ij}h_{ij}}{E_{ij}h_{ij}+E_{io}h_{io}}\epsilon_{ij} \quad (11)$$

Through moment balance analysis of $PZT_{ij}$ and utilizing the constitutive relationships of Eq. (11) that couple $\epsilon_{io}$ and $\epsilon_{ij}$ with the temporal voltage signal $V_{ij}(t)$, one obtains:

$$M_{ij}(y,t) = \frac{d_{31}E_{io}h_{io}E_{ij}(h_{ij}+h_{io})}{2(E_{io}h_{io}+E_{ij}h_{ij})^2} \times [E_{io}h_{io}a_{ij}+E_{ij}h_{ij}a_{io}]V_{ij}(t) \quad (12)$$

$$M_{ij}(y,t) = P_{yij}\cdot V_{ij}(t) \quad (13)$$

Thus, $U_{pij}$ can be written as:

$$U_{pij} = -\frac{1}{2}\int_{A_{ij}}\left[P_{yij}\cdot V_{ij}(t)\left(\frac{\partial^2 u}{\partial y^2}\right)\right]dx\,dy \quad (14)$$

The total strain energy associated with the PZT-generated moments can be expressed as:

$$U_p = -\frac{1}{2}\sum_{i=1}^{n}\int_{A_{ij}}\left[P_{yij}\cdot V_{ij}(t)\left(\frac{\partial^2 u}{\partial y^2}\right)\right]dx\,dy \quad (15)$$

Incorporating the $U_p$ into the Lagrangian equation, one can obtain $F_c$, the control force created by the $PZT_{ij}$ actuators on the $i^{th}$ beam:

$$\vec{F}_c = \frac{1}{2}\sum_{j=1}^{m}\frac{V_{ij}}{a_{ij}b_{ij}}\left[P_{yij}\vec{C}(x_{ij},y_{ij})\right] \quad (16)$$

$$\vec{C} = \int_{\xi_j-\frac{b_j}{2b_o}}^{\xi_j+\frac{b_j}{2b_o}} \Psi''d\xi \quad (17)$$

where $m$ is the number of PZT actuators on the appendage $i$.

## IV. CONTROL MODELS

In this section, two innovative control techniques are developed and applied to the vibrational dynamics model of a satellite with PZT-mounted beam-type appendages. The initial method employed is Nonlinear Model Predictive Control (NMPC), which leverages a predictive model of the system's dynamics to determine optimal control actions over a limited future timeframe. The second technique is an AI-based Nonlinear AutoRegressive with eXogenous inputs (NARX) control method, which operates without requiring an explicit mathematical model of the system dynamics.

### A. NPMC Model

This section presents the Nonlinear Model Predictive Control (NMPC) model for the coupled structural and vibrational control of PZT-mounted beam-type satellite appendages. The conventional MPC formulation for this system is given by (Kalaycioglu and de Ruiter, 2023):

$$\min \int_0^{T_p} \left( (\vec{y}(t) - \vec{y}_r(t))^T K_y (\vec{y}(t) - \vec{y}_r(t)) + \vec{S}^T(t) K_s \vec{S}(t) \right) dt \quad (18)$$

subject to:

$$\dot{\vec{y}} = \vec{g}(\vec{y}) + L\vec{S}, \quad \text{and} \quad \vec{z} = \vec{g}_z(\vec{y}) + H\vec{S} \quad (19)$$

$$\vec{y}(0) = \vec{y}(t_0) \quad (20)$$

$$\vec{S}_{\min} \leq \Lambda \vec{S} \leq \vec{S}_{\max} \quad (21)$$

where the prediction horizon is denoted by $T_p$, and the weighting matrices $K_y$ and $K_s$ are positive definite. The functions $\vec{g}(\vec{y})$, $\vec{g}_z(\vec{y})$, $L$, and $H$ originate from the nonlinear system equations in Eqs. (6) and (16). The matrix $\Lambda$ serves as a positive definite scaling matrix, used to adjust for the saturation limits.

Upon discretization of the system equations, one can obtain the following difference equations:

$$\vec{y}(k+1) = \hat{A}(\hat{g}(k))\vec{y}(k) + \hat{B}(\hat{g}(k))\vec{S}(k) \quad (22)$$

$$\vec{z}(k) = \hat{C}(\hat{g}_z(k))\vec{y}(k) + \hat{D}(\hat{g}(k))\vec{S}(k) \quad (23)$$

$$\hat{g}(k) = f_g(\vec{y}(k)) \quad (24)$$

$$\vec{S}_{\min} \leq \Lambda \vec{S}(k) \leq \vec{S}_{\max} \quad (25)$$

At the discrete time step $k$, the output vector $\vec{z}(k)$ is obtained from measurements taken at that specific moment. The matrices in Eqs. (22-25) are provided in Kalaycioglu and de Ruiter (2023).

At every sampling point, the NMPC algorithm calculates the discrete state variables $\vec{y}(k)$ and control signals $S(k)$ by optimizing the following cost function:

$$C_n = \frac{1}{2} \sum_{j=1}^{N_p} \Big( (\vec{y}(k+j) - \vec{y}_r(k+j))^T \\ \times K_y (\vec{y}(k+j) - \vec{y}_r(k+j)) \\ + \vec{S}(k+j-1)^T K_s \vec{S}(k+j-1) \Big) \quad (26)$$

subject to:

$$\vec{y}(k+j+1) = \hat{A}(\hat{g}(k+j))\vec{y}(k+j) + \hat{B}(\hat{g}(k+j))\vec{S}(k+j) \quad (27)$$

$$\vec{z}(k+j) = \hat{C}(\hat{g}(k+j))\vec{y}(k+j) + \hat{D}(\hat{g}(k+j))\vec{S}(k+j) \quad (28)$$

$$\vec{S}_{\min} \leq \Lambda \vec{S}(k+j) \leq \vec{S}_{\max} \quad (29)$$

However, when constraints are active, the optimal control solution cannot simply be derived by setting the derivative of the cost function to zero. Instead, the optimality conditions must be derived based on the *Karush-Kuhn-Tucker (KKT) conditions*, which take into consideration the constraints on the control inputs.

To solve the optimization problem and satisfy the KKT conditions, the MATLAB Quadratic Programming (QP) solver is used. The QP solver automatically incorporates the constraints from Eq. (21), ensuring that the control inputs respect the bounds and minimize the cost function subject to these constraints.

The QP solver minimizes the following quadratic cost function:

$$C_n = \frac{1}{2} \sum_{j=1}^{N_p} \Big[ \vec{S}(k+j)^T \hat{P} \vec{S}(k+j) + 2(\vec{g})^T \vec{S}(k+j) \Big] \quad (30)$$

where:

$$\hat{P} = (\hat{B}^T K_y \hat{B} + K_s) \quad (31)$$

$$(\vec{g})^T = \left( \vec{y}(k+j)^T \hat{A}^T K_y \hat{B} - \vec{y}_r(k+j)^T K_s \hat{B} \right) \quad (32)$$

By using the QP solver, the control inputs are guaranteed to satisfy the KKT conditions, ensuring that the solution is optimal while adhering to the constraints.

The result is a robust NMPC formulation that effectively controls the structural and vibrational dynamics of satellite appendages equipped with PZT actuators, achieving the desired performance while respecting input constraints.

### B. NARX-Based Control for Vibration Suppression of PZT Mounted Beam-Type Satellite Appendages

The NARX (Nonlinear AutoRegressive with eXogenous inputs) model is an AI-based method for system identification and control, particularly suitable for complex, nonlinear systems where traditional modeling approaches may be inadequate (He, 2015 and Song, 2023). Unlike NMPC, the NARX model does not require an express analytical model of the system, making it highly versatile and adaptive.

In the context of vibration suppression for PZT-mounted beam-type satellite appendages, the NARX model is employed to estimate the system states and predict the required control actions. The control inputs, consisting of the PZT control forces are determined based on the predicted system outputs.

NARX neural networks offer a promising framework for controlling systems characterized by high nonlinearity and the need for adaptability. NARX-based control utilizes the network's capacity for inverse dynamics control, directly learning the inverse dynamics model to achieve a unity transfer function between the actual and desired output.
One of the significant advantages of using NARX-based control is its adaptive nature. The model can be updated online as new data becomes available, allowing it to adjust to changes in system dynamics and external disturbances. This functionality proves to be especially advantageous in space missions, where unpredictable conditions frequently arise (Roghanchi, 2019).

The adaptive control strategy can be described as follows:
   (i) Continuously collect real-time data from the satellite's sensors.
   (ii) Periodically retrain the NARX model with the latest data to capture any changes in the system's behavior.
   (iii) Update the control inputs based on the updated model predictions.

The proposed approach augments the standard NARX architecture with a neural network discriminator (NND) (Song, 2023). This NND serves as an online learning mechanism, continuously refining the NARX model's connection weights by discriminating between input-output pairs obtained in real time. This dynamic adaptation significantly enhances the controller's robustness to external disturbances and system parameter variations, addressing a key limitation of traditional direct inverse control.

By embedding the NND within the control strategy, the NARX-based system maintains high performance and precision even under unpredictable conditions. This makes it particularly well-suited for applications where adaptability and robustness are critical, such as spacecraft and robotics in complex dynamic environments. The key advantages include the following:
   (i) Nonlinearity: NARX models inherently capture complex nonlinear relationships between system inputs and outputs.
   (ii) Adaptability: The NND enables online learning, ensuring continuous model refinement in response to changing conditions.
   (iii) Robustness: The dynamic nature of the control strategy provides resilience to external perturbations and parameter drift.

*Model Formulation*: The NARX model predicts the future states of a system based on past inputs and outputs. The general form of the NARX model can be expressed as:

$$y(t+1) = h\big(y(t), y(t-1), \ldots, y(t-d_y), \\ S(t), S(t-1), \ldots, S(t-d_s)\big) \quad (33)$$

where $y(t)$ is the output at time $t$, $S(t)$ is the input at time $t$, $d_y$ is the maximum lag in the output, and $d_s$ is the maximum lag in the input.

The mathematical formulation of the NARX-based control can be expressed as:

$$\vec{y}(t+1) = \bar{A}\vec{y}(t) + \bar{B}\vec{S}(t) \quad (34)$$

$$\vec{z}(t) = \bar{C}\vec{y}(t) + \bar{D}\vec{S}(t) \quad (35)$$

where $\bar{A}$, $\bar{B}$, $\bar{C}$, and $\bar{D}$ are matrices obtained from the trained NARX model.

*Implementation:* The NARX-based control is implemented as follows:
   (i) Collect historical data of the system outputs $\vec{y}(t)$ and inputs $S(t)$.
   (ii) Train the NARX model using this data to capture the system dynamics.
   (iii) Use the trained NARX model to predict future states $\vec{y}(t+1)$ based on current and past states and inputs.
   (iv) Compute the control inputs $S(t)$ required to achieve the desired system performance.

*Training and Validation:* The NARX model can be trained using either actual historical data or simulated data from a developed dynamics model. This training process optimizes the model's weights and biases to minimize the prediction error between the model's output and the actual observed output. Commonly employed algorithms for this optimization include Levenberg-Marquardt backpropagation and Bayesian regularization, which are known to enhance the model's robustness and generalization capabilities (Rahrooh, 2009; Yan, 2016 and Kelley, 2024).

The training process can be summarized as follows:
   (i) Collect a comprehensive dataset of system inputs $S(t)$ and outputs $\vec{y}(t)$.
   (ii) Preprocess the data to remove noise and normalize values.
   (iii) Initialize the NARX network with a suitable architecture, typically including several hidden layers and neurons.
   (iv) Train the network using a portion of the dataset, employing either the Levenberg-Marquardt algorithm or Bayesian regularization.
   (v) Assess the performance of the trained model on a distinct validation dataset to verify its ability to generalize effectively to new, unseen data.

V. RESULTS AND DISCUSSION

The proposed NMPC and NARX-Inverse Dynamics based control approach were evaluated through MATLAB/Simulink simulations. Figure 3 illustrates the control block diagram.

Performance of the two methods and that of the PD control were compared under various disturbance conditions, demonstrating effective vibration suppression in the flexible appendages. Notably, NMPC performance was dependent on model accuracy, while the NARX approach proved more effective under model uncertainties and disturbances. Key metrics included vibration amplitude reduction, response time, and stability under varying conditions.

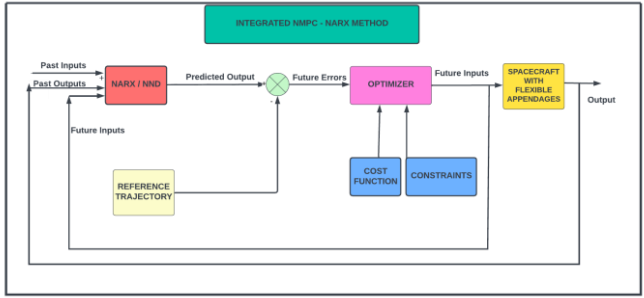

Fig. 3. NMPC-NARX Control Method

A comprehensive case study was performed on a satellite equipped with beam-type appendages mounted with PZTs. The NMPC method utilized the developed structural dynamics model, while the NARX model was trained on historical data and PZT voltage-induced structural motions. The trained NARX model was subsequently employed and constantly updated with the support of the neural network discriminator (NND). This NND served as an online learning mechanism to predict the control inputs necessary for vibration suppression and attitude maintenance while the system was subject to disturbances.

Simulation results showed that both NMPC and NARX based control methods outperformed a traditional control strategy like PD control, providing faster response times and better vibration suppression in the case of existence of significance dynamics model uncertainties and disturbances. These findings demonstrated the potential of NARX models for advanced vibration control in space applications.

*A. Simulation Setup, Network Configuration and Training*

The simulation setup involved creating a detailed structural dynamics model of the satellite with PZT-mounted beam-type appendages in MATLAB/Simulink. The NARX and NND models were also integrated into this simulation environment to evaluate its performance under various conditions.

The NARX neural network was configured with input delays ranging from 1 to 2 time steps and feedback delays also spanning from 1 to 2 time steps. This configuration allows the network to utilize both recent and slightly older input and output values to predict future outputs, capturing the temporal dynamics of the system effectively. The network architecture included a single hidden layer consisting of 10 neurons, utilizing ReLU as the activation function. This selection was made to achieve an optimal balance between the complexity of the model and the efficiency of computation.

The network was trained using the Levenberg-Marquardt algorithm, an optimization method known for combining the strengths of both the Gauss-Newton and gradient descent approaches. This algorithm enhances convergence speed and improves the network's ability to reach an optimal solution, making it particularly effective for neural network training.

The dataset was split into three distinct subsets: training, validation, and testing. The training subset was used to update the network's weights and biases, while the validation subset monitored the training process to prevent overfitting by assessing the network's performance after each epoch. Finally, the test subset was reserved for evaluating the network's generalization performance on unseen data once the training was completed.

The performance of the network was evaluated using the Mean Squared Error (MSE) metric, which calculates the average of the squared differences between predicted and actual values. This metric offers a clear assessment of the network's accuracy, where lower MSE values reflect higher prediction precision and better overall performance.

The main parameters for the simulation were a sampling time of 0.01 seconds, a prediction horizon spanning 10 steps, and a control horizon of 5 steps. The disturbance scenarios examined included random vibrations, sudden impacts, and persistent external forces.

To comprehensively assess the NMPC and NARX-based control, a range of disturbance scenarios were considered. These included random vibrations to simulate low-magnitude disturbances and evaluate the control's ability to manage minor perturbations, sudden impacts to assess responsiveness and stability under high-magnitude forces, and continuous external forces to evaluate long-term stability maintenance.

The available measurements included sensor data on vibration amplitudes and system response times, which were crucial in evaluating the control strategies. Measurement noise was explicitly modeled in the simulation to reflect real-world conditions, with Gaussian noise added to the vibration and impact sensor data, characterized by a standard deviation of 0.05 and a mean of zero. The disturbances were also carefully modeled: random vibrations were generated using a white noise process with a standard deviation of 0.1 and a bandwidth of 5 Hz, while sudden impacts were simulated as impulse forces with magnitudes ranging from 5 N to 20 N. Continuous external forces were applied as sinusoidal functions with frequencies ranging from 0.5 Hz to 2 Hz and amplitudes from 0.2 N to 1 N to test long-term stability. The NARX neural network inherently accounted for system dynamics and provided accurate predictions based on the input-output data.

Performance was evaluated using key metrics such as vibration amplitude reduction, response time, and stability.

The detailed parameters of the PZT sensors and actuators, spacecraft, and the flexible beam-type appendages utilized in the simulations are listed in Table I.

| Parameter | Values | Unit |
|---|---|---|
| beam Dimensions $(a_{io}, b_{io}, h_{io})$ | $(0.3, 1, 2 \times 10^{-3})$ | m |
| PZT Dimensions $(a_{ij}, b_{ij}, h_{ij})$ | $(0.1, 0.1, 0.5 \times 10^{-3})$ | m |
| Young's Modulus of Elasticity beam and PZT $E_{io}, E_{ij}$ | $6.9 \times 10^{10}$ | Pa |
| The Locations of PZTs $(x_{ij}, y_{ij})$ | $[(0.1, 0.1), (0.5, 0.1), (0.9, 0.1)]$ | m |
| Poisson's Ratio | 0.33 | none |
| Density of the beam $\rho$ | $2.7 \times 10^3$ | kg/m$^3$ |
| PZT Electric Coefficient $d_{31}$ | $-1.75 \times 10^{-10}$ | V$^{-1}$ |
| Moments of Inertia of Satellite | $\begin{bmatrix} 1.2 & -0.05 & -0.03 \\ -0.05 & 3.02 & -0.02 \\ -0.03 & -0.02 & 8.02 \end{bmatrix}$ | kg·m$^2$ |

TABLE I
PHYSICAL PARAMETERS FOR THE SATELLITE AND THE APPENDAGES AND PZT USED IN MATLAB/SIMULINK SIMULATIONS

The elastic deformation of the beam was discretized and expanded in terms of modal shape functions. The shape functions for the beam were obtained from the eigenfunctions of a clamped-free (cantilever) beam. Figure 4 illustrates the mode shapes for a clamped-free beam.

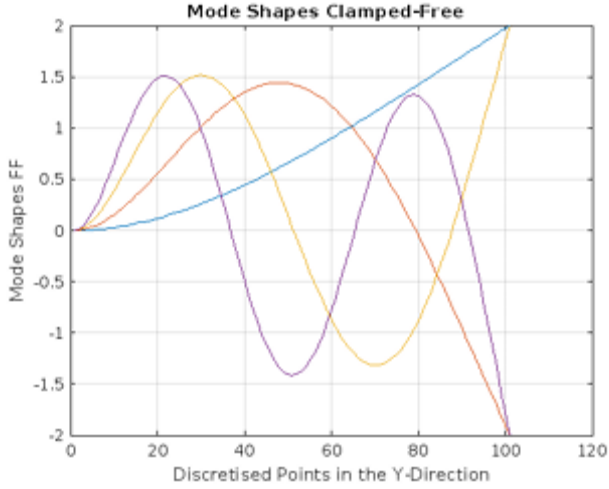

Fig. 4. Eigenshapes for a Clamped-Free Beam.

*A.    Results of the NMPC Method*

Figure 5 demonstrates the time histories of the first three modal time coefficient functions for both NMPC and PD control methods. Across all three cases, the NMPC method exhibited superior performance in suppressing vibrations compared to the PD control. Specifically, the modal time coefficient functions under NMPC displayed faster decay rates and smaller amplitudes, indicating more rapid and effective vibration suppression.

The visual representation clearly shows that the NMPC method consistently outperforms the PD control across all modes, highlighting its robustness and reliability in various operational conditions. These characteristics suggest that the NMPC method is more efficient at damping out oscillations, thereby stabilizing the system more quickly.

Figure 6 presents the time variation of the PZT modal force, $F_c$, for both NMPC and PD control methods. The NMPC method produced smoother and more controlled PZT forces compared to the PD control, which exhibited higher frequency oscillations. This observation suggested that NMPC can achieve better vibration suppression with less aggressive control actions, potentially leading to reduced energy consumption and less wear on the PZT actuators. The smoother force profile under NMPC indicated that the actuators were subjected to less stress, which could enhance their longevity and reduce maintenance requirements.

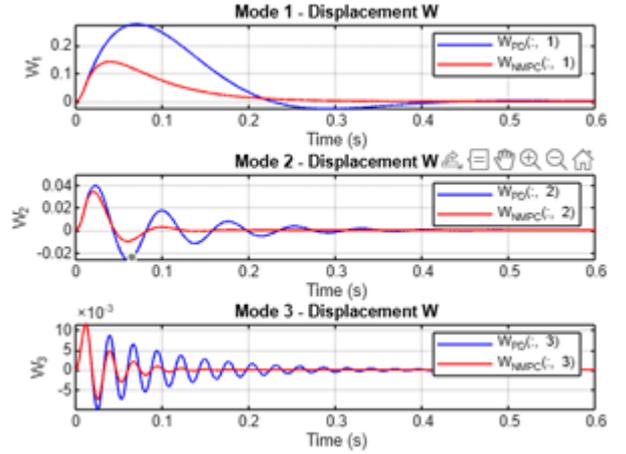

Fig. 5. Comparison of Modal Time Coefficient Functions for NMPC and PD control methods.

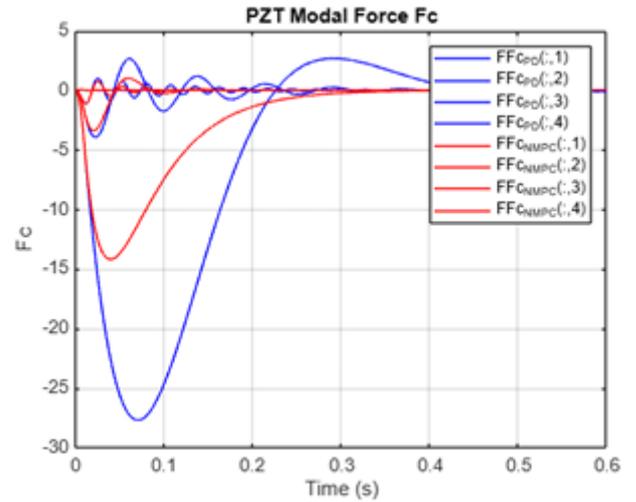

Fig. 6 Time variation of PZT Modal Force Fc under NMPC and PD control.

*B.    Results of the NARX Method*

Figure 7 compares the time-varying modal coefficients for both NARX and a traditional PD control under conditions of model uncertainty and disturbances. The NARX method produced superior results compared to the PD control, which exhibited higher frequency oscillations. This observation suggested that NARX can achieve better vibration suppression with less aggressive control actions.

Figure 8 presents the time variation of the PZT modal force, $F_c$, for both the NARX and PD control methods. The NARX method produced superior results compared to the PD control, which exhibited higher frequency oscillations. This observation suggested that NARX can achieve better vibration suppression with less aggressive control actions.

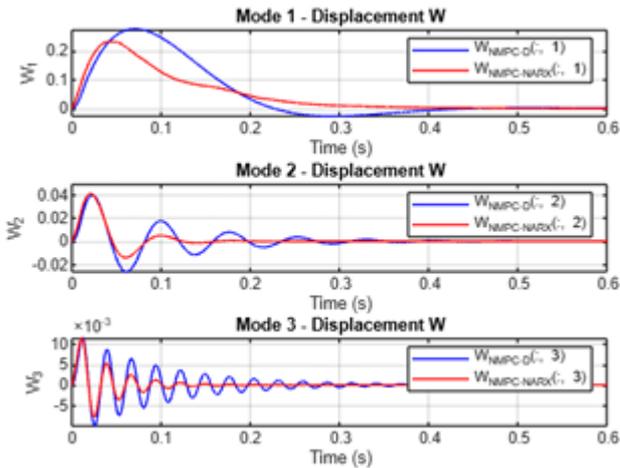

Fig. 7. Comparison of Modal Time Coefficient Functions for NMPC and PD control methods.

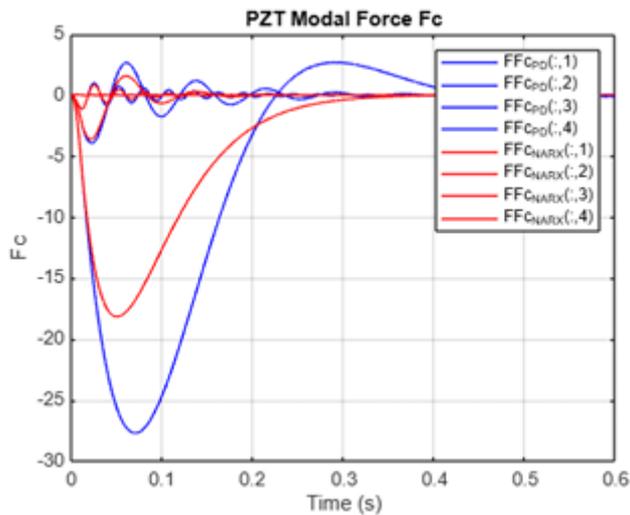

## VI. CONCLUSION

This study introduced a novel approach to vibration control of satellite beam antennas utilizing Piezoelectric (PZT) sensors and actuators integrated with Nonlinear Model Predictive Control (NMPC) and NARX control techniques. By developing a structural dynamics model tailored for flexible beam-type antennas, this study provided a robust framework for understanding and managing the complex interactions between satellite attitude and structural vibrations.

The NMPC method effectively predicted future system behaviors and optimized control actions, demonstrating its capability for precise and robust control of flexible structures. The NARX-based control, augmented with a neural network discriminator (NND), exhibited superior adaptability to dynamic environments and system parameter variations, addressing a key limitation of traditional model based control. This continuous refinement of the NARX model ensured high performance and precision even under unpredictable conditions, making it particularly well-suited for spacecraft and robotic applications.

The NARX-based control approach demonstrated its capability to capture and manage intricate, nonlinear dynamics effectively, offering a compelling alternative to conventional control techniques that typically require explicit mathematical modeling. Its successful application in simulations underscored its potential for enhancing the stability and performance of satellite systems.

Key challenges encountered in developing real-time vibration control systems included accurately modeling the dynamic response of large elastic structures, managing nonlinearities and uncertainties, and ensuring computational efficiency within the limited resources of a spacecraft. The integration of sensors and actuators for precise data collection and effective control also proved crucial. Addressing these challenges remains essential for advancing the field and ensuring the robustness of control systems in the harsh environment of space.

This research underscored the potential of integrating advanced control strategies with smart materials to improve the performance and stability of flexible satellite structures. The findings highlighted the potential of combining NMPC and NARX techniques with PZT actuators to significantly improve satellite stability and performance. This work contributed to the advancement of space technology by providing effective solutions for controlling flexible satellite appendages, a critical aspect for future space missions.